\documentclass{jltp}
\usepackage{graphicx,bm}
\title{Magnetic flux penetration and motion in antiferromagnetic superconductors}
\author{Tomasz Krzyszto\'n}
\address{Institute of Low Temperature and Structure Research,\\
Polish Academy of Sciences,\\ 50-950 Wroc\l{}aw, P.O.Box 1410,\\
Poland} \runninghead{\bfseries T. Krzyszto\'n}{\bfseries
Antiferromagnetic superconductors}
\begin{document}
\maketitle
\begin{abstract}
The paper reviews a concept of induced spin-flop domain inside
vortices in an antiferromagnetic superconductor. Such phenomenon
may occur when an external magnetic field is strong enough to flip
over magnetic moments in the core of the vortex from their ground
state configuration. The formation of the domain structure inside
vortices modifies the surface energy barrier of the
superconductor. During this process the entrance of the flux is
stopped and a newly created state exhibits perfect shielding. Such
behavior should be visible as a plateau on the dependence of flux
density as a function of the external magnetic field. The end of
the plateau determines the critical field, which has been called
the second critical field for flux penetration. Moreover, it is
predicted and described how this phenomenon modifies flux creep in
layered superconductors. The various scenarios of changing the
creep regime from thermal to quantum and vice versa at constant
temperature are discussed. PACS numbers: 74.60.-w, 74.25.Ha.
\end{abstract}
\section{INTRODUCTION}
The discoveries of ternary Rare Earth (RE)
Chevrel Phases REMo$_{6}$S$_{8}$ and RERh$_{4}$B$_{4}$
\cite{Ternary} compounds with regular distribution of localized
magnetic moments of RE atoms have proved conclusively the
coexistence of various types of magnetism with superconductivity.
Intensive experimental and theoretical works have shown that 4f
electrons of RE atoms responsible for magnetism and 4d electrons
of molybdenum chalcogenide or rhodium boride clusters responsible
for superconductivity are spatially separated and therefore their
interaction is weak. In many of these systems superconductivity
coexists rather easily with antiferromagnetic order, where usually
the Neel temperature $T_{N}$ is lower than the critical
temperature for superconductivity $T_{c}$. On the other hand,
ferromagnetism and superconductivity cannot coexist in bulk
samples with realistic parameters. Quite often the ferromagnetic
order is transformed into a spiral or domain-like structure,\cite{Kasperczyk80}
depending on the type and strength of magnetic anisotropy in the
system.\cite{BulBuzdKulPanj,Maple95} For almost two decades the
problem of the interaction between magnetism and superconductivity
has been overshadowed by high temperature superconductivity (HTS)
found in copper oxides. However, the recent discovery of the
presence of magnetic order in Ru-based superconductors
\cite{Bauer,Pringle,KlamutX,Houzet} has triggered a new series of
experiments and inspired a return to the so-called coexistence
phenomenon.\cite{Houzet} Most recently, the interplay between
magnetism and superconductivity was studied in d-electron
UGe$_{2}$\cite{Saxena} and ZrZn$_{2}$\cite{Pfleiderer}, where
itinerant ferromagnetism may coexist with superconductivity, and
in heavy fermion UPd$_{2}$Al$_{3}$,\cite{Sato} where magnetic
excitons are present in the superconducting phase. Among classic
magnetic superconductors, the Chevrel phases have been studied
most intensively. These compounds are mainly polycrystalline
materials. However, some specific features can be measured only on
single crystals. One such effect is a two-step flux penetration
process, predicted in Ref.(\onlinecite{Krzy80,Krzy84}) and later
observed solely in the antiferromagnetic superconductor (bct)
ErRh$_{4}$B$_{4}$.\cite{Muto86} This very interesting phenomenon
was recently rediscovered in DyMo$_{6}$S$_{8}$,\cite{Rogacki2001}
although good quality single crystals of the classic
antiferromagnetic superconductors have been a long time available
and measured.

The DyMo$_{6}$S$_{8}$ compound with $T_{c}=1.6~\rm {K}$ exhibits
transition from the paramagnetic to the antiferromagnetic state at
$T_{N}=0.4~\rm {K}$. Its crystal structure can be described as
interconnected Mo$_{6}$S$_{8}$ units and Dy ions. One such unit is
a slightly deformed cube where S atoms sit at the corners and Mo
atoms are situated at the cube-faces. The Mo$_{6}$S$_{8}$ units
are arranged in a simple rhombohedral lattice and Dy ions are
located in the center of the unit cell. The magnetic moments of Dy
ions form a simple structure consisting of $(100)$ planes with
moments of $8.7~\mu _{B}$ alternately parallel and antiparallel to
the $[111]$ rhombohedral axis. Neutron experiments performed on
DyMo$_{6}$S$_{8}$ in an applied magnetic field at $T=0.2~\rm {K}$
have revealed in the intensity spectrum a number of peaks
characteristic for ferromagnetic order.\cite{Thomlinson79} These
peaks begin to develop at $H_{0}=200~\rm {Oe}$, much below the
superconducting upper critical field $H_{c2}$. Thus, in
DyMo$_{6}$S$_{8}$ a kind of ferromagnetic order coexists with
superconductivity in the same manner as antiferromagnetism. For a
field applied parallel to the $[111]$ direction (magnetic
easy-axis direction), the ferromagnetic order is a spin-flop
type.\cite{Thomlinson82}

\begin{figure}
\centerline{\includegraphics[height=2.5in]{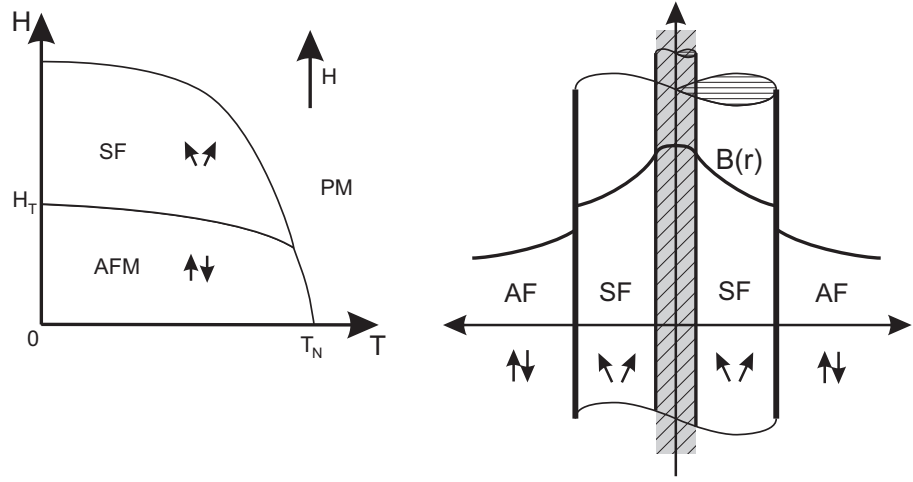}}
\caption{On the left panel: phase diagram of two-sublattices
uniaxial antiferromagnet. The external field is directed along the
anisotropy axis. On the right panel: the magnetic structure of an
Abrikosov vortex. Gray area corresponds to the vortex core where
the spin-flop transition originates. This model has also been used
by other authors \cite{Buzdin,Wong}.} \label{fig1}
\end{figure}

This feature is easy to understand. Consider the well known phase
diagram of a two-sublattice antiferromagnet. An infinitesimal
magnetic field applied perpendicular to the easy axis makes the
ground antiferromagnetic configuration unstable against the phase
transformation to the canted phase. On the contrary, if the
magnetic field is applied parallel to the easy axis the
antiferromagnetic (AF) phase is stable up to the thermodynamic
critical field $H_{T}$ as is seen on Fig.\ref{fig1} (left panel).
When the field is further increased, a spin-flop (SF) phase
develops in the system. Let us assume that in the
antiferromagnetic superconductor the lower critical field fulfills
the relation $H_{c1}<\frac{1}{2}H_{T}$ and that the external field
$H_{0}$ is applied parallel to the easy axis. When
$H_{c1}<H_{0}<\frac{1}{2}H_{T}$ the Abrikosov vortices appear
entirely in the AF phase. When $H_{0}$ is increased beyond
$\frac{1}{2}H_{T}$ the phase transition to the SF phase originates
in the core, because near $H_{c1}$ the field intensity in the core
is approximately twice $H_{c1}$.\cite{Tinkham} The spatial
distribution of the field across the vortex is a function
decreasing from the center as is shown in Fig.~\ref{fig1} (right
panel). Thus, the magnetic field intensity outside the core is
less then $H_{T}$ and, therefore, the rest of the vortex remains
in the AF phase. The radius of a spin-flop domain grows as the
field is increased. The formation of domains inside the vortices
should be accompanied by the modification of the surface energy
barrier.\cite{Krzy84,Krzy2002}
\begin{figure}
\centerline{\includegraphics[height=2.35in]{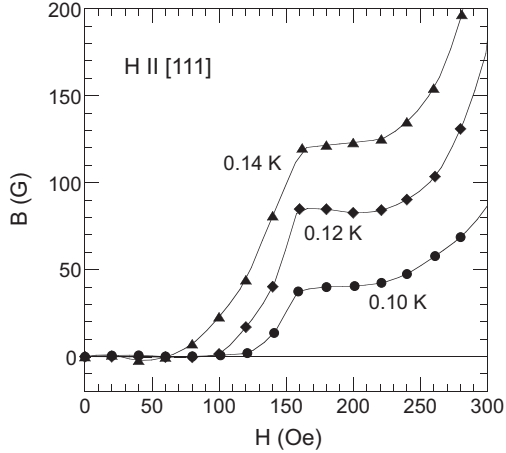}}
\caption{Magnetic induction for DyMo$_6$S$_8$ single crystal in
the virgin state measured as a function of an applied field for
three temperatures below $T_N = 0.4~\mathrm{K}$. The field
direction is oriented parallel to the magnetic easy axis of the
crystal. Each $B(H)$ curve exhibits characteristic plateau
indicating that a number of vortices is kept constant when the
external field is increased. The calculated values of the second
critical field for flux penetration $H_{en2}(B)$ are in very good
agreement with the measured ones.\cite{Krzy2002}} \label{fig2}
\end{figure}
This process leads to a state of the superconductor in which flux
entrance is temporarily prohibited, flux density in the sample is
constant as the applied field increases (see Fig.~\ref{fig2}). In
order to kill this state the external field should be increased
above certain second critical field for flux penetration
$H_{en2}(B)$. Then, the vortices penetrating the sample will have
the spin flop domains created along the cores.

The above considerations apply to the classical superconducting
Chevrel phases as well as to the high $T_{c}$ superconductors. The
present work is inspired by the above described discovery and the
hope that the same behavior could possibly be observed in some of
the layered superconducting structures. Indeed, the situation
seems to be very similar in layered HTS. Here magnetic order is
produced by the regular lattice of RE ions occupying isolating
layers electrically isolated from the superconducting Cu-O planes.
Therefore spin interaction between the local magnetic moments and
the conduction electrons is to weak to inhibit superconductivity.
The typical example of the layered system is ErBa$_2$Cu$_3$O$_7$.
This compound has tetragonal unit cell with small orthorhombic
distortion in the a-b plane.\cite{Lynn92,Zaretsky} The $Er$ ions
form two sublattice antiferromagnetic structure of magnetic
moments lying parallel and antiparallel to the $b$ direction in
the $ab$ plane.\cite{Lynn90} Recently discovered RE nickel
boride-carbides\cite{Muller} may serve as an another example of
layered magnetic superconductors. The layered structure of RE
nickel boride-carbides is reminiscent of that of HTS and  consists
of RE-carbon layers separated by Ni$_{2}$B$_{2}$ sheets.
\cite{Sinha95,Szymczak95,Eskildsen2001,Canfield98} For example in
ErNi$_2$B$_2$C the antiferromagnetic structure is associated with
magnetic moments of Er$^{+3}$ ions, which order below 6 K in a
transversely polarized planar sinusoidal structure propagating
along $a$ or $b$ axis with Er moments parallel to the $a$ or $b$
axis respectively.
\section{LONDON THEORY}
\begin{figure}
\centerline{\includegraphics[height=2.2in]{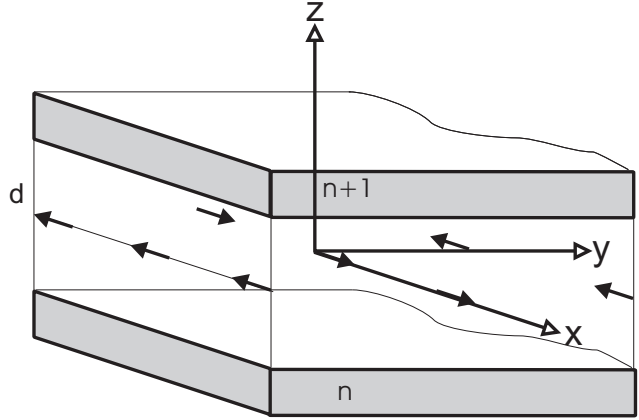}}
\caption{Schematic drawing of a piece of the layered
superconductor. The shaded areas (n,n+1) represent superconducting
layers. The bold arrows represent magnetic moments of RE ions
lying in the isolating layers. The axes of the reference frame are
shown.} \label{fig3}
\end{figure}

In the following we consider the structure shown on
Fig.~\ref{fig3} that we believe simulates a real structure of many
antiferromagnetic layered superconductors. A good candidate to
show the above described phenomenon should possess the isolating
layers with the magnetic moments of RE ions running parallel and
antiparallel to the direction (easy axis) lying in the $ab$ plane.

We start description of our problem in terms of the
Lawrence-Doniach energy functional. In this approach a layered
superconductor is described by the superconducting planes with the
interlayer distance d, as shown on Fig.~\ref{fig3}. The
antiferromagnetic subsystem consisting of RE ions is confined to
the isolating layers. The magnetic moments are running parallel
and antiparallel to the x-axis (easy axis). The Lawrence-Doniach
functional is obtained from the standard Ginzburg-Landau  energy
by discretization of the kinetic energy in the z-direction.
\begin{eqnarray}
F_{S}=\int {d^2}r d \sum_{n} \Bigg[\frac{\hbar ^{2}}{2m} \left|
\left( -i\mathbf{\nabla_{(2)}}+\frac{2ie}{\hbar
}\mathbf{A_{(2)}}\right)\Psi_n \right|^{2} + \mathsf{a} \left| \Psi_n
\right|^{2}+\frac{1}{2} \mathsf{b} \left| \Psi_n \right|^{4}+ \nonumber \\
+\frac{\hbar^{2}}{2\mathcal{M}d^{2}}\left|\Psi_{n+1} \exp
\left(\frac{2ei}{\hbar} \int_{nd}^{(n+1)d} A_z dz \right)-\Psi_n
\right|^2 \Bigg] \label{eq1}
\end{eqnarray}
The quantity $e,m,$ denote the charge of the free
electron and the mass of the current carrier in the $ab$ plane,
whereas $\mathcal{M}$ denotes the mass of the current carrier in
the z-direction;  $\mathbf{A} = (A_{(2)},A_z)$ denotes the
vector potential. The antiferromagnetic two sublattices subsystem
with single ion anisotropy is  described with the following energy
density functional
\begin{equation}
f_{M}=\sum_{n}
\Big\{J\mathbf{M}_{1n}\cdot\mathbf{M}_{2n}+K\sum\limits_{i=1}^{2}\left(
M_{in}^{x}\right)^{2}-\left| \gamma
\right|\sum\limits_{i=1}^{2}\sum\limits_{j=x,y,z}
(\mathbf{\nabla}M_{in}^{j})^{2}\Big\} \label{eq2}
\end{equation}
where $\mathbf{M}_{n}=\mathbf{M}_{1n}+\mathbf{M}_{2n}$ is the sum
of the magnetization vectors of the sublattices in the n-th
insulating layer, $M_{in}^{x}$ is the component of the
magnetization sublattice vector along the anisotropy axis in the
n-th layer, $J$ denotes the exchange constant between two
sublattices, $K$ is the single ion anisotropy constant,
$\sqrt{\left| \gamma \right|}$ is the magnetic stiffness length,
and\ $M_0=\left| \mathbf{M}_{1n}\right| =\left|
\mathbf{M}_{2n}\right| $. The components of the total
magnetization vector $\mathbf{M}$ have the following form in both
sublattices:$M_{iy}=M_{0}\sin\theta_{i},~M_{ix}=M_{0}\cos\theta_{i}$,
where $\theta _{i}$  (canted spin angle) is the angle between the
magnetization in the sublattice and the external magnetic field
directed along the $x$-axis. The AF $(\theta _{1}=0,\theta
_{2}=\pi )$ and SF phases $(\theta_{1}=-\theta _{2}=\theta )$ are
in thermodynamic equilibrium in an applied field equal to the
thermodynamic critical field
\begin{equation}
H_{T}=M_{0}[K(J-K)]^{1/2}.  \label{eq3}
\end{equation}
The canted spin angle of the SF phase is then expressed as
\begin{equation}
\cos \theta=\frac {KM_{0}}{H_{T}}. \label{eq4}
\end{equation}
Finally we add the magnetic field energy to obtain the free energy
of the entire system
\begin{equation}
F=F_{s}+ \int \Big\{ f_{M}+\frac{\mu _{0}}{2} (\mathbf{b}-\mathbf{M})^{2}
\Big\} dV. \label{eq5}
\end{equation}
According to experiments the antiferromagnetic order is very weak
affected by the presence of superconductivity, then it is
reasonable to neglect the effect of superconductivity on the
exchange interaction in $F$. Instead we introduce electromagnetic
coupling between the magnetic subsystem and superconducting current $\mathbf{j}_s$. This
means that both order parameters $\Psi_n $ and $\mathbf{M}$ are
coupled through the vector potential $\mathbf{A}$
\begin{equation}
\mathbf{b}=\rm{rot}\mathbf{A} \label{eq6}
\end{equation}
\begin{equation}
\mu_0\mathbf{j}_s=\rm{rot}(\mathbf{b}-\mathbf{M}), \label{eq7}
\end{equation}
where $\mathbf{b}\ $ is the vector of the microscopic magnetic field.

The functional~(\ref{eq5}) can be treated in the London
approximation by assuming a constant modulus $\Psi_n$ within the
planes and allowing only for the phase degree of freedom. The
equilibrium conditions of the system are the result of
minimization of the Gibbs free energy functional
$G=F-\displaystyle\int\mathbf{b}(\mathbf{b}-\mathbf{M})dV$. The London
equations resulting from~(\ref{eq5}) are following:
\begin{eqnarray}
&&b_{x}+\lambda _{c}^{2}\frac{\partial }{\partial y}{\rm rot}_{z}
(\mathbf{b}-\mathbf{M})-\lambda_{ab}^{2}\frac{\partial }{\partial z}{\rm
rot}_{y}(\mathbf{b}-\mathbf{M})=\varphi_{0}\delta (y)\delta (z) \nonumber \\
&&b_{y}+\lambda_{ab}^{2}\frac{\partial }{\partial z}{\rm rot}_{x}
(\mathbf{b}-\mathbf{M})-\lambda _{c}^{2}\frac{\partial }
{\partial x}{\rm rot}_{z}(\mathbf{b}-\mathbf{M})=0 \nonumber \\
&&b_{z}+\lambda _{ab}^{2}\frac{\partial }{\partial x}{\rm rot}_{y}
(\mathbf{b}-\mathbf{M})-\lambda _{ab}^{2}\frac{\partial }{\partial y}{\rm
rot}_{x}(\mathbf{b}-\mathbf{M})=0   \label{eq8}
\end{eqnarray}
$\lambda_{ab}=\lambda_c\sqrt{\mathcal{M}/m}$,
$r_j=d\sqrt{\mathcal{M}/m}$. The London model of continuous
superconductors may be used at length scales larger than the
coherence length $\xi$, i.e. the core dimension.
\begin{figure}
\centerline{\includegraphics[height=1.6in]{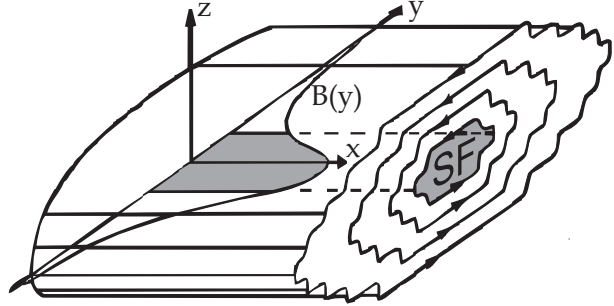}}
\caption{Single Josephson vortex lying in the $ab$ plane along the
$\hat{x}$-axis. The SF domain induced along the phase core is
shown in the gray area.} \label{fig4}
\end{figure}
The structure of a vortex lying in the $ab$ plane in a layered
superconductor with Josephson coupling between adjacent layers
resembles the Abrikosov`s one except that the order parameter does
not vanish anywhere.\cite{Clem90} Instead there exists a region,
$r_j$ along the plane and $d$ perpendicular to it, where the
Josephson current $j_{z}$ is of the order of the critical current.
In this region, which plays the role of the vortex core, the
London model fails. Away from the core the streamlines of the
shielding supercurrents, which also represents contours of
constant magnetic field, are elliptical except for the zigzags due
to the intervening insulating layers (see Fig.~\ref{fig4}).

The equations (\ref{eq8}) should be supplemented by the
appropriate set of differential equations describing the spatial
distribution of magnetization. Simple conjecture can make the
calculations less complex and at the same time does not
oversimplify the problem. We assume \cite{Krzy94} that the
magnetic moment is constant across the magnetic domain
\begin{equation}
\left| \bf{M}\right| =\left\{
\begin{array}{ccc}
M & if & \rho <\rho _{m} \\
0 & if & \rho >\rho _{m}
\end{array}
\right. \label{eq9}
\end{equation}
where $\ \rho _{m}$ is the dimensionless radius of the magnetic
domain in the coordinate system of the elliptical cylinder
($x=x,y=\lambda _{c}\rho \cos \varphi ,z=\lambda _{ab}\rho \sin
\varphi $). Then the solution of Eq.~(\ref{eq8})~in the
cylindrical reference frame for a single Josephson vortex is
given by the modified Bessel functions $K_{0}$ and $I_0$
\begin{eqnarray}
b_\mathrm{ SF} &=&  C_{1}K_{0}\left(\rho\right)
+C_{2}I_{0}\left(\rho\right)~~for~~\rho_j
<\rho\leq \rho_{m}  \nonumber \\
b_\mathrm{ AF} &=& C_{3}K_{0}\left(\rho\right)~~for~~\rho>\rho_{m}
\label{eq10}
\end{eqnarray}
($\rho_j $ denotes the dimensionless phase coherence length ) with
the following boundary conditions:
\begin{eqnarray}
b_{\mathrm{SF}}\left(\rho_m\right) &=& \mu_{0}H_{T}+M = B_{T}\nonumber\\
b_{\mathrm{AF}}\left(\rho_{m}\right) &=&\mu_{0} H_{T} \label{eq11}
\end{eqnarray}
These conditions, together with the flux quantization condition,
are used to calculate the arbitrary constants in (\ref{eq10}).
\begin{eqnarray}
&&C_{1}=\frac{\displaystyle B_{T} \rho_m I_{1}\left(
\rho_m\right)-\left[\mu_0 H_{T}\rho_m\frac{K_{1}\left( \rho_m
\right)} {K_{0}\left( \rho_m\right)}-\frac{\varphi _{0}}{2\pi
\lambda_c\lambda_{ab}}\right] I_{0}\left(
\rho_m\right)}{\displaystyle\rho_{m} K_{1} \left( \rho_m\right)
I_{0}\left( \rho_m\right) -I_{0}
\left( \rho_m\right) +\rho_m K_{0}\left( \rho_m\right)I_{1}\left( \rho_m\right)}\nonumber \\
&&C_{2}=\frac{\displaystyle{B_{T}\left[ \rho_m K_{1}\left( \rho_m
\right) -1\right] +\left[\mu_0 H_{T}\rho_m\frac{K_{1}\left( \rho_m
\right)}{K_{0}\left( \rho_m\right)}-\frac{\varphi
_{0}}{2\pi\lambda_c\lambda_{ab}}\right] K_{0}\left(
\rho_m\right)}}{\displaystyle\rho_m K_{1}\left( \rho_m\right)
I_{0}\left( \rho_m\right) -
I_{0}\left(\rho_m\right) +\rho_m K_{0}\left(\rho_m\right)I_{1}\left(\rho_m\right)}\nonumber \\
&&C_{3}=\frac{\mu_0 H_{T}}{K_{0}\left(\displaystyle \rho_m\right)}
\label{eq12}
\end{eqnarray}
Finally we write free energy of the isolated vortex
\begin{eqnarray}
\varepsilon &=&\frac{\lambda_c\lambda_{ab}}{2\mu_0}\oint_{\sigma
_{1}}d \bm{\sigma}
\left\{\left[\textbf{b}_\mathrm{SF}\left(\bm{r}\right)-\textbf{M}\right]\times
\mathrm{rot}
\textbf{b}_\mathrm{ SF}\left(\bm{r}\right)\right\}\nonumber\\
&+&\frac{\lambda_c\lambda_{ab}}{2\mu_0}\oint_{\sigma_{2}}d
\bm{\sigma}\left\{ \textbf{b}_\mathrm{AF}\left(\bm{r}\right)\times
\mathrm{rot}\textbf{b}_\mathrm{AF}\left(\bm{r}\right)\right\}
\label{eq13}
\end{eqnarray}
where
$\bm{r}=\left(\displaystyle{\frac{y}{\lambda_{ab}},\frac{z}{\lambda_c}}\right)$
is the position of the vortex line, $\sigma_1$ denotes the surface
of the phase core, and $\sigma_2$ the surface of the SF domain
respectively. The integrals in (\ref{eq13}) performed as line
integrals along the contours of the cross sections of the
appropriate  surfaces  give $\varepsilon _{1}$ - the line tension
of the vortex. The minimum of $\varepsilon _{1}$ with respect to
$\rho_m$ determines
\begin{equation}
\rho_{m }^{2}=\frac{5\phi _{0}}{8\pi\lambda_c\lambda _{ab}B_{T}}
\label{eq14}
\end{equation}
\subsection{Equilibrium energy of the vortex lattice}
The London equations can be rewritten for the lattice of vortices
in the following way:
\begin{equation}
\mathbf{B}+\mathrm{rot}\mathrm{rot}\mathbf{B}=\frac{\phi
_{0}}{\lambda_c\lambda _{ab}}\sum_m\delta(\bm{r}-\bm{r}_m)
\label{eq15}
\end{equation}
where $\mathbf{r}_m$ specify the positions of the phase cores of the
vortices. The solution of Eq.~(\ref{eq15}) is then a superposition
\[\mathbf{B}(\bm{r})=\sum_m \mathbf{b}_m(\bm{r}-\bm{r}_m)\]
of the solutions $\mathbf{b}_m(\bm{r}-\bm{r}_m)$ of isolated
vortices at the points $\bm{r}_m$. The free energy of the system
can thus be  written as
\begin{equation}
F=\frac{\lambda_c\lambda_{ab}}{2\mu_0}\oint_{\sigma }d
\bm{\sigma}(\mathbf{B}\times\mathrm{rot}\mathbf{B}) \label{eq16}
\end{equation}
The above symbolic surface integral is taken over the surfaces of
the phase cores and the surfaces of the SF domains. The energy of
the Meissner state is chosen as zero of the energy scale. Again,
when the surface integrals are replaced by contour ones we get
line energy of the system. This, in turn, multiplied by vortex
density $n$ gives $f$ the free energy density of the system. After
some transformations one can derive the following formula
\begin{equation}
f=n\varepsilon _{1} + n\phi_0 H_T (\ln \beta)^{-1} \sum_m
K_0(r_m)\ \ ; \ \
\beta=\sqrt{\frac{\displaystyle\pi\lambda_c\lambda_{ab}
B_T}{\displaystyle\phi _{0}}} \label{eq17}
\end{equation}
here the sum is over all vortices excluding the one in the origin,
and $r_m$ denotes the distance of a vortex from the origin. The
lattice sum may now be replaced by integral in the $yz$-plane over
a smoothed vortex density, excluding the area $n^{-1}$ associated
with the single flux line in the origin. The free energy density
then reduces to
\begin{equation}
f=n\varepsilon _{1}+B^2\Big(\frac{\displaystyle H_T}{\displaystyle
B_T}\Big)\Big(\frac{\displaystyle\beta}{\displaystyle\ln\beta}\Big)+B
\frac{\displaystyle{H_T}}{\displaystyle
4\ln\beta}\sqrt{\frac{\displaystyle 4\lambda_{ab}}
{\displaystyle27\lambda_c}}\ln\Big(\frac{\displaystyle{c}}{\displaystyle
\sqrt{\lambda_c\lambda_{ab}}} \Big) \label{eq18}
\end{equation}
\[ \Big(\frac{\displaystyle c}{\displaystyle\sqrt{\lambda_c\lambda_{ab}}}\Big)^2=
\frac{\displaystyle 1}{\displaystyle\beta^2}\Big(\frac{
\displaystyle B_T} {\displaystyle B}\Big)\sqrt{\frac{\displaystyle
4\lambda_{ab}}{\displaystyle 27\lambda_c}},\] here $c=|\bm{c}_1|$
denotes the length of the basal vector of the nonequilateral
triangular unit cell, and $2|\bm{c}_2|=c\sqrt{1+\tan^2\alpha}$ (
$\alpha $ is the angle between both vectors),~
$\tan\alpha=\sqrt{\frac{3\lambda_c}{\lambda_{ab}}}$.\cite{Kogan81}
To determine the equilibrium state it is necessary to minimize the
Gibbs free energy density with respect to magnetic induction. The
result yields an implicit equation for the constitutive relation between magnetic induction and thermodynamic magnetic field.
\begin{equation}
H-\frac{\varepsilon _{1}}{\phi_0}=B\Big(\frac{\displaystyle
H_T}{\displaystyle B_T}\Big)\Big(\frac{\displaystyle 2
\beta}{\displaystyle\ln\beta}\Big)+\frac{\displaystyle{H_T}}{\displaystyle
4\ln\beta}\sqrt{\frac{\displaystyle 4\lambda_{ab}}
{\displaystyle27\lambda_c}}\ln\Big(\frac{\displaystyle{c}}{\displaystyle
\sqrt{\lambda_c\lambda_{ab}}} \Big) \label{eq19}
\end{equation}
\section{TWO STEP FLUX PENETRATION}
Consider a semi-infinite specimen in the half space $y\geq 0$, the
vortex and the external magnetic field $H_0$ running parallel to the
surface in the $x$ direction. The presence of a surface of the
superconductor leads to a distortion of the field and current of
any vortex located within a distance of the order of penetration
depth from the surface. To fulfill the requirement that the
currents cannot flow across the surface of the superconductor we
need to introduce an image vortex with  the vorticity opposite to
the real one. Both vortices, direct and image, interact as real
ones except that the interaction is attractive. In the low flux
density regime Clem \cite{ClemLT} has shown that there exist two
regions: a vortex-free region of the width $y_{ff}$ near the
surface of the sample, and a constant flux density region for
$y>y_{ff}$. Within the vortex-free area one can introduce the
locally averaged magnetic field $B_{M}$ which is  a linear
superposition of the Meissner screening field, the averaged direct
vortices flux density exponentially decreasing towards the surface
from its interior value $B$ at $y=y_{ff}$, and averaged image
vortices flux density. In our problem the $x$ component of this
superposition can be approximated by
\begin{equation}
B_{M}=B\cosh \left(\frac{y_{ff}-y}{\lambda_{ab}}\right)
\label{eq20}
\end{equation}
The boundary condition $B_M(0)=\mu_0 H_0$ determines the thickness
of the vortex-free region
\begin{equation}
y_{ff}=\lambda_{ab} \cosh^{-1}
\left(\frac{\mu_0H_{0}}{B}\right)\label{eq21}
\end{equation}
We assume that the test vortex line is lying within vortex free
region at a point
$\bm{r}=\left(\displaystyle{\frac{y}{\lambda_{ab}},0}\right)$, and
its image at
$\bm{r}=\left(\displaystyle{-\frac{y}{\lambda_{ab}},0}\right)$
outside the superconductor. Now the local field of the test vortex
can be understood as a superposition of the following fields
\begin{eqnarray}
\mathbf{B}_\mathrm{ SF}&=&\mathbf{b}_\mathrm{
SF}\left(\bm{r}\right)-\mathbf{b}_\mathrm{ AF}
\left(2\bm{r}\right) +\hat{x}B_{M}\left(\bm{r}_{ff}-\bm{r}\right) \nonumber \\
\mathbf{B}_\mathrm{ AF}&=&\mathbf{b}_\mathrm{
AF}\left(\bm{r}\right)-\mathbf{b}_\mathrm{ AF}\left(2\bm{r}\right)
+\hat{x}B_{M}\left(\bm{r}_{ff}-\bm{r}\right)\label{eq22}
\end{eqnarray}
where
$\bm{r}_{ff}=\left(\displaystyle{\frac{y_{ff}}{\lambda_{ab}},0}\right)$,
and $\hat{x}$ denotes the unit vector in the $x$ direction. Having
determined the local magnetic field we can write the Gibbs free
energy of the test vortex line as
\begin{eqnarray}
G&=&\frac{\lambda_c\lambda_{ab} }{2\mu_0}\oint_{\sigma _{1}}d
\bm{\sigma}\left\{\left[ \mathbf{B}_\mathrm{
SF}\left(\bm{r}\right)-2\mu_0\mathbf{H}_{0}-\mathbf{M}\right]
\times\mathrm{rot} \mathbf{B}_\mathrm{SF}\left(\bm{r}\right)\right\}\nonumber\\
&+&\frac{\lambda_c\lambda_{ab} }{2\mu_0}\oint_{\sigma _{2}}d
\bm{\sigma} \left\{\left[\mathbf{B}_\mathrm{
AF}\left(\bm{r}\right)-2\mu_0\mathbf{H}_{0}\right]
\times\mathrm{rot} \mathbf{B}_\mathrm{AF}\left(\bm{r}\right)\right\}\nonumber \\
&+&\frac{\lambda_c\lambda_{ab} }{2\mu_0}\oint_{\sigma _{2}}d
\bm{\sigma}
\left\{\hat{x}B_{M}\left(\bm{r}_{ff}-\bm{r}\right)\times\mathrm{rot}
\mathbf{B}_\mathrm{AF}\left(\bm{r}\right)\right\} \label{eq23}
\end{eqnarray}
After some transformations \cite{Krzy94,ClemLT} one can obtain
the Gibbs free energy per unit length $\mathcal{G}$
\begin{equation}
\mathcal{G}=\mathcal{G}_{1}+\mathcal{G}'_{1}+\mathcal{G_M}
\label{eq24}
\end{equation}
where
\begin{eqnarray}
\mathcal{G}_{1} &=& \varepsilon
_{1}-\frac{\lambda_c\lambda_{ab}\pi}{4\mu_0}D_1b_\mathrm{ AF}
\left(2r\right)\nonumber\\
\mathcal{G}'_{1} &=&
-\frac{\lambda_c\lambda_{ab}\pi}{2\mu_0}D_1\left[b_\mathrm{ AF}
\left(r_{ff}\right)-b_\mathrm{ AF}\left(r_{ff}+r\right)\right]\nonumber\\
\mathcal{G_M} &=& -\frac{\lambda_c\lambda_{ab}\pi}{2\mu_0}\left[
D_{1}\mu_{0} H_{0} - D_{2} B_{M}\left(r_{ff}-r\right)\right]
\label{eq25}
\end{eqnarray}
and
\begin{eqnarray}
D_{1}&=& \left. -\rho_j\frac{d b_\mathrm{
SF}(\rho)}{d\rho}\right|_{\rho=\rho_j} \left. -\rho_m\frac{d
b_\mathrm{ SF}(\rho)}{d\rho}\right|_{\rho=\rho_m}\left.
-\rho_m\frac{d b_\mathrm{ AF}(\rho)}{d\rho}\right|_{\rho=\rho_m}\nonumber\\
D_{2}&=&\left. -\rho_j\frac{d b_\mathrm{
SF}(\rho)}{d\rho}\right|_{\rho=\rho_j}\left. -\rho_m\frac{d
b_\mathrm{ SF}(\rho)}{d\rho}\right|_{\rho=\rho_m}\left.
-2\rho_m\frac{d b_\mathrm{ AF}(\rho)}{d\rho}\right|_{\rho=\rho_m}
\label{eq26}
\end{eqnarray}
$\mathcal{G}_{1}$ describes the interaction of the test vortex
with its image, $\mathcal{G}'_{1}$ is a correction term introduced
by Clem,\cite{ClemLT} and $\mathcal{G_M}$ describes the
interaction energy of the test vortex with the modified Meissner
field. To find the conditions of the vortex entrance and exit, one
has to solve a force balance equation for the test vortex, at the
surface of the sample, and at the edge of the flux-filled area,
respectively. A calculation using $\mathcal{G}_{1}$ and
$\mathcal{G_M}$ alone gives non vanishing force on the test vortex
at $\bm{r}=\bm{r}_{ff}$. However, the force should be zero there,
because $\mathcal{G_M}$ is supposed to account for all the image
vortices. To avoid double counting the image vortex one can
subtract from the self-energy  a contribution of the excess image
fixed at $\bm{r}=-\bm{r}_{ff}$. One can easily check that
$\mathcal{G}'_{1}$ is negligible at the surface of the sample and
has no influence on the conditions of the flux entrance. When the
flux starts to enter the sample, $H_0=H_{en2}(B)$,
\begin{equation}
y_{ff}=y_{en}=\lambda_{ab} \cosh^{-1}\Big(\frac{\mu_0
H_{en2}(B)}{B}\Big) \label{eq27}
\end{equation}
and the energy barrier is moved toward the surface within
$\rho_m$. Thus, one can derive from the force balance equation
\begin{equation}
-\frac{D_1}{2D_2}\left.\frac{d b_\mathrm{
AF}(\rho)}{d\rho}\right|_{\rho=\rho_m}=B\sinh\Big(\frac{y_{en}}{\lambda_{ab}}\Big)
\label{eq28}
\end{equation}
The left hand side of the above equation gives $H_{en2}(0)=
H_T\beta(2\ln\beta)^{-1}$. This field may be thought as the second
critical field for flux penetration calculated in the single
vortex approximation \cite{Krzy94}. Combining Eqs.~(\ref{eq27})
and~(\ref{eq28}) one can finally obtain
\begin{equation}
H_{en2}(B)=\sqrt{B^2+\Big(\frac{\mu_0
H_{T}\beta}{2\ln\beta}\Big)^2} \label{eq29}
\end{equation}
In the opposite case, when the flux exits the sample, the surface
energy barrier tends to the edge of the flux-filled zone. Similar
considerations as the above show that
\begin{equation}
\mu_0 H_{ex2}(B)\simeq B \label{eq30}
\end{equation}
The measure of the height of the energy barrier against flux
entrance is
\[\Delta H_{en}(B)=\left|H_{en2}(B)-H_{eq}(B)\right|,\]
and against flux exit
\[\Delta H_{ex}(B)=\left|H_{eq}(B)-H_{ex2}(B)\right|,\]
where $H_{eq}$ is given by Eq.~(\ref{eq19}).
\begin{figure}
\centerline{\includegraphics[height=2.2in]{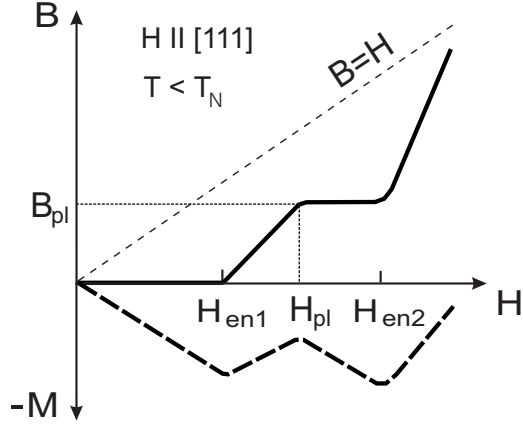}}
\caption{Schematic drawing of the magnetization process. $H_{en1}$
denotes the first penetration field for vortices without magnetic
structure. $H_{pl}$ is the applied field which originates SF
transitions inside vortices, and $B_{pl}$ is the corresponding
flux density.  $H_{en2}$ is the entrance field for the vortices
possessing magnetic structure. The lower dashed line shows $M(H)$
dependence. The region of plateau on $B(H)$ corresponds to the
region of the second negative slope on $M(H)$.} \label{fig5}
\end{figure}

Let us make a short summary of the calculations and visualize the
results on schematic magnetization curve shown in the
Fig.~\ref{fig5}. When the external field is not strong enough to
create the SF domains inside vortices, than the magnetization
process of the sample being entirely in the AF phase is as
follows. The vortices without magnetic structure start to enter
the specimen at $H_{en1}$. When the field is increased up to the
value $H_{pl}$, which is of the order of $H_T$, the SF domains are
created. Now, the screening current must redistribute its flow in
order to keep constant the flux carried by the vortex. This
feature is easily seen from Eqs.~(\ref{eq10})-(\ref{eq12}). The
redistribution of the screening current changes the surface energy
barrier preventing vortices from entering the sample as expressed
in (\ref{eq28}). It means that the density of vortices $n$ is
kept constant. Consequently the averaged flux density in the
sample $B=n\varphi_0$ remains constant when the external field is
increased. In Fig.~\ref{fig5} this feature is visible as a plateau
on the $B(H)$ curve, or alternatively as a second negative slope
on the $M(H)$ curve. The vortices start to penetrate the sample
when the external field reaches the right edge of the plateau. We
call this value, given by (\ref{eq29}), second critical field
for flux penetration $H_{en2}$.

To find the thermodynamic critical field $H_{T}$, and then to
calculate $H_{en2}(B)$ the following argumentation is used. At low
fields, in the vicinity of the lower critical field $H_{c1}$, the
intensity of the field in the vortex core is
$2H_{c1}$.\cite{Clem90} When the external field is increased the
field intensity in the vortex core increases because of the
superposition of the fields of the surrounding vortices. The field
intensity in the core must reach $H_T$ in order to originate a
transition to the SF phase. Thus, taking into account only the
nearest neighbors we can write for the nonunilateral triangular
lattice
\begin{equation}
H_{T}=2H_{c1}+z\frac{\varphi _{0}}{\pi
\lambda_c\lambda_{ab}\mu_0}\left[K_{0}\left(\frac{c}{\lambda_{ab}}\right)+
2K_{0}\left(\frac{c}{2\lambda_{ab}}\sqrt{\frac{3\lambda_c}{\lambda_{ab}}}\right)\right]
\label{eq31}
\end{equation}
here $c$ corresponds to the value $B_{pl}$ of the flux density for
which the penetration process stops, see Fig.~\ref{fig5}. From
the relation $B_{pl} =
2\varphi_{o}\sqrt{\lambda_{ab}}/(c^{2}\sqrt{3\lambda_c})$ one can
compute $c$, which in turn may be inserted back into
Eq.~(\ref{eq31}). It is easy to estimate the saturation
magnetization $M_0$ taking into account the volume of the
elementary cell. Then, equations~(\ref{eq3}) and  (\ref{eq4}) can be
used to calculate $M$ in the SF-phase domain
\begin{equation}
M = 2M_{0}\cos\theta = \frac{2KM_{0}^{2}}{H_{T}} \label{eq32}
\end{equation}
\section{INTRINSIC PINNING}
\begin{figure}[!htb]
\centerline{\includegraphics[height=2.1in]{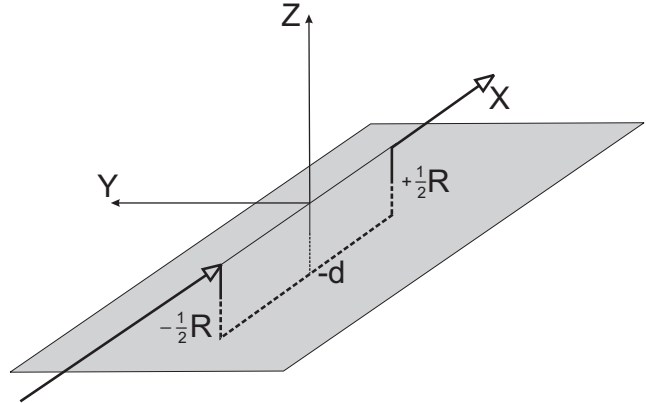}}
\caption{The double-kink excitation form a critical nucleus. The
part of the vortex is thrown to the adjacent layer.} \label{fig6}
\end{figure}

The first quantitative approach toward intrinsic pinning in
layered superconductors was based on the observation that the
superconducting order parameter should have a periodic spatial
variation across the layers. For the present considerations,
however, the method of critical nucleus developed in
\cite{Chakravarty90} is much more convenient. The activated
nucleus consists of a kink-antikink excitation, that is, a vortex
line segment is thrown to the adjacent layer, thereby creating two
pancake vortices of opposite vorticity, as shown in the
Fig.~\ref{fig6}.
The activation energy can be regarded as the energy barrier for
intrinsic pinning. Depending on the magnitude of the driving
current density the process may continue as the single vortex
activation or the activation of the vortex bundle.
\subsection{Thermal activation of vortices}
First, consider the activation of a segment of a vortex to the
neighboring interlayer spacing (see Fig.~\ref{fig6}). The energy
associated with this process can be written as:
\begin{equation}
U=\delta E+V_{K,-K}(R)-(j-j_{0})\varphi _{0}dR. \label{eq33}
\end{equation}
The subscript ''a'' or ''b'' of $U$, $\delta E$ and $j_{0}$
indicates that this quantity is calculated for $a$ or $b$
direction in the plane. $\delta E $ is the amount of condensation
and magnetic domain energy that is lost at two points of the layer
threaded by the kinks separated by a distance $R$. $V_{K,-K}(R)$
is the kink-antikink interaction energy. The term proportional to
the driving current $j$ is due to the Lorentz force. The term
proportional to $j_{0}$ is the energy associated with the
distortion of the line due to the formation of nucleus. This term
can be estimated from the simple considerations
\[
j_{0}\varphi _{0}d\sim \frac{1}{2}\int dydzC(y,z)\left(
\frac{\partial u_{z} }{\partial z}\right) ^{2}
\]
where the Fourier transform of the compression modulus is given by
~\cite{Brandt92}
\[
C(k_{y},k_{z})=\frac{B^{2}}{\mu _{0}(1+\lambda
_{ab}^{2}k_{z}^{2}+\lambda _{c}^{2}k_{y}^{2})}
\]
By taking $dydz\sim \varphi _{0}/B,u_{z}\sim d,\frac{\partial
}{\partial z} \sim k_{z}\sim k_{y}(\lambda _{c}/\lambda _{ab})$
the integral can be estimated as follows
\begin{equation}
j_{0b}=\frac{Bd}{4\lambda _{ab}^{2}}. \label{eq34}
\end{equation}
However, in the $a$ direction, according to (\ref{eq14}) we
have an additional contribution from the magnetic domain $dydz\sim
5\varphi _{0}/8B_{T}$. Thus we get
\begin{equation}
j_{0a}=j_{0b}+\frac{5dB_{T}}{128\lambda _{ab}^{2}}. \label{eq35}
\end{equation}
As the current density $j$ drops below $j_{0}$ a single-vortex
line can no longer be activated due to the confinement energy
provided by the vortex lattice. The current $j_0$ plays very
important role in all above calculations. For the following values
$d\sim 10^{-9}m$, $\xi_c\sim 3\times 10^{-10}m$, $\xi _{ab}\sim
3\times 10^{-9}m$, $r_j\sim 10^{-8}m$ and their typical
temperature dependence we can estimate
\[
\frac{j_{0b}}{j_{GL}}\sim \frac{Bd\xi _{ab}}{\varphi _0}\sim
10^{-3}B\sqrt{ \frac{T_c}{T_c-T}}
\]
where $j_{GL}=4\epsilon _{0}/(\varphi _{0}\xi _{ab}3\sqrt{3})$ is
the Ginzburg-Landau critical current density and $\epsilon
_{0}=\varphi _{0}^{2}/(16\pi ^{2}\mu _{0}\lambda _{ab}^{2})$.
Since the depairing current is of the order of $\ 10^{13}\left[
T_c/(T_c-T)\right] ^{3/2}\ $ then the current\ $j_{0b}$ is of the
order of $10^{10}B \left[ T_c/(T_c-T)\right]$. Although there are
no precise measurements of SF transition in the AF HTS, we assume
that $\mu _0H_T\sim 40mT$. The typical value of $ 5.5\mu_B$ per RE
atom per unit cell gives $M\sim 0.37T$. It is possible now to
estimate the change of $j_0$ due to the creation of SF domain
along the vortex :
\[
\frac{j_{0b}}{j_{0a}}\sim 1+0.625\frac{B_T}B\sim 3
\]
providing that\ $H_T/H_{c1}-1<<1$. The energy $\delta E$ is
calculated from Eq.~(\ref{eq8}) with the right-hand sides
representing the vortex cores:\cite{Krzy98}
\[\left\{ \left| x\right| >\frac R2,y=0,z=0\right\}\left\{ x=\pm
\frac R2,y=0,0<z<-d\right\}\left\{ \left| x\right| <\frac
R2,y=0,z=-d\right\}\]
The solution is then substituted to the free energy functional
(\ref{eq5}). Taking the limit $R\rightarrow \infty $ we
exclude the energy of the kink-antikink interaction. Because the
calculations are involved we write down only the results.
\begin{equation}
\delta E_{b}=2d\epsilon _{0}\ln \frac{r_{j}}{\xi_{ab}}\label{eq36}
\end{equation}

\begin{equation}
\delta E_{a}=d\epsilon _{a}\ln \frac{r_{j}}{\xi_{ab}} \label{eq37}
\end{equation}
where $\epsilon _{a}=\frac{77}{64}\epsilon _{0}\ln \left[ \varphi
_{0}/\left(\pi r_{j}^{2}B_{T}\right)\right]$. The energy of
kink-antikink interaction was calculated in\cite{Chakravarty90}
\begin{equation}
V_{K,-K}(R)=-\frac{d^{2}\epsilon _{0}}{2\lambda
_{ab}}f\left(\frac{R}{\lambda _{c} }\right)\label{eq38}
\end{equation}
where
\[
f\left(\frac{R}{\lambda _{c}}\right)=\left\{
\begin{array}{ccc}
\left(\lambda _{c}/R\right)-\ln (r_{j}/\xi _{ab}) & for &
\begin{array}{cc}
r_{j}<<R<< & \lambda _{c}
\end{array}
\\
2\left(\lambda _{c}/R\right)^{3}\exp \left( -R/\lambda _{c}\right)
& for & R>>\lambda _{c}
\end{array}
\right.
\]
If we introduce the quantity
$I_{a,b}=2(j-j_{0a,b})/(j_{GL}3\sqrt{3})$, then Eq.~(\ref{eq33})
can be rewritten in the following way
\begin{eqnarray}
U_{b} &=&2d\epsilon _{0}\left\{ \ln \frac{r_{j}}{\xi
_{ab}}+I_{b}\frac{R}{ \xi _{ab}}-\frac{d}{4\lambda
_{ab}}f\left(\frac{R}{\lambda
_{c}}\right)\right\} \nonumber \\
U_{a} &=&d\left\{ \epsilon _{a}\ln \frac{r_{j}}{\xi
_{ab}}+2\epsilon _{0}I_{a}\frac{R}{\xi _{ab}}-\frac{d\epsilon
_{0}}{2\lambda _{ab}} f\left(\frac{R}{\lambda _{c}}\right)\right\}
\label{eq39}
\end{eqnarray}
The critical size of the nucleus $R_{c}$ is given as a minimum of
Eq.~(\ref{eq39}) with respect to $R$. In the approximation
$r_{j}<<R<<\lambda _{c}$ corresponding to the current regime $\xi
_{c}d/\lambda _{ab}^{2}<<I_{a,b}<<\xi _{c}/d $ we get

\begin{eqnarray}
R_{ca,b}^{2} &=&\xi _{ab}^{2}\frac{d}{4I_{a,b}\xi _{c}}  \nonumber \\
U_{b}^{c} &=&2d\epsilon _{0}\left\{ \ln \left( \frac{r_{j}}{\xi
_{ab}}\right) -\sqrt{\frac{dI_{b}}{\xi _{c}}}\right\}  \\
U_{a}^{c} &=&d\left\{ \epsilon _{a}\ln \left( \frac{r_{j}}{\xi
_{ab}}\right) -2\epsilon _{0}\sqrt{\frac{dI_{a}}{\xi_{c}}}\right\}
\nonumber \label{eq40}
\end{eqnarray}
For the opposite case $ R>>\lambda _{c}$ and $ \xi _{c}d/\lambda
_{ab}^{2}>>I_{a,b}$
\begin{eqnarray}
R_{ca,b} &=&\lambda _{c}\ln \left( \frac{d\xi _{c}}{I_{a,b}\lambda
_{ab}^{2}}\right)  \nonumber \\
U_{b}^{c} &=&2d\epsilon _{0}\left\{ \ln \left( \frac{r_{j}}{\xi
_{ab}}\right) -\frac{\lambda _{ab}I_{b}}{\xi _{c}}\ln \left(
\frac{d\xi _{c}}{I_{b}\lambda _{ab}^{2}}\right) \right\}  \\
U_{a}^{c} &=&d\left\{ \epsilon _{a}\ln \left( \frac{r_{j}}{\xi
_{ab}}\right) -\frac{\lambda _{ab}I_{a}}{\xi _{c}}\ln \left(
\frac{d\xi _{c}}{I_{a}\lambda _{ab}^{2}}\right) \right\} \nonumber
\label{eq41}
\end{eqnarray}
When the driving current drops below $j_{0}$ the critical nucleus
is 3D object. In our case it is a parallelepiped of the height $R$
along the bundle and of the section $S$ across it. The activation
energy is a sum of the volume energy due to the Lorentz force and
the surface energy.
\begin{equation}
U_{a,b}=-jBdRS+\delta E_{a,b}\left( \frac{BS}{\varphi _{0}}\right)
+j_{0a,b}dR\sqrt{BS\varphi _{0}} \label{eq42}
\end{equation}
The second term is the loss of condensation energy (and magnetic
domain energy in the case of $a$ direction) on both surfaces
perpendicular to the bundle multiplied by the number of vortices
threading these surfaces. The third term is the elastic energy
released in the surface parallel to the shifted vortex
$j_{0a,b}dR\varphi _{0}$ multiplied by the number of shifted
vortices $\sqrt{BS/\varphi _{0}}$ (one vortex per plane). The
critical nucleus is then $S_{c}=(\varphi _{0}/B)(j_{0a,b}/j)^{2}$
, $R_{c}=\delta E_{a,b}/(jd\varphi _{0})$, and the activation
energy is
\begin{equation}
U_{a,b}^{c}=\delta E_{a,b}\left( \frac{j_{0a,b}}{j}\right) ^{2}.
\label{eq43}
\end{equation}
\subsection{Thermal creep}
The resistive mechanism in the mixed state is determined by the
activation process leading to magnetic flux motion (creep). This
motion induces electric field which can be observed on the
current-voltage characteristic. We consider the motion of
activated kinks along the layers of the length $L$ along the
magnetic field direction. Assume that each double-kink can reach
the boundary of the sample before the new one is created. The mean
electric field associated with this motion is given by
\begin{equation}
E=BPLdS_{c} \label{eq44}
\end{equation}
where $P$ is the activation probability per unit volume and unit
time. For thermal activation this probability is given by $\
P\sim \exp \left( -U_{c}/k_{B}T\right) \ $. There is however a
crossover temperature $T_{0}$ below which quantum tunneling of
vortices is dominating. The probability for quantum tunneling is
finite even for $T=0$. The Neel temperature for layered
antiferromagnetic superconductors varies from hundreds of mK (0.6K
for ErBa$_{2}$Cu$_{3}$O$_{7} $) to several Kelvin (6.8K for
ErNi$_{2}$B$_{2}$C) and therefore both mechanisms of activation
are present in these compounds. The preexponential factors and
$T_{0}$ cannot be calculated in the framework of thermodynamic
considerations alone. Fortunately, it was shown in \cite{Tekiel}
that the activation probability of macroscopic quantum excitations
is proportional to $j^{3}$. Thus we can assume that $P=\alpha
_{0}j^{3}\exp \left( -U_{c}/k_{B}T\right) $. Now we can calculate
the current-voltage characteristics for the current density
$j<<j_{0}$
\begin{eqnarray}
E_{a} &=&\varphi _{0}dL\alpha _{0}j_{0a}^{2}j\exp \left\{
-\frac{\delta E_{a}
}{k_{B}T}\left( \frac{j_{0a}}{j}\right) ^{2}\right\}  \\
E_{b} &=&\varphi _{0}dL\alpha _{0}j_{0b}^{2}j\exp \left\{
-\frac{\delta E_{b} }{k_{B}T}\left( \frac{j_{0b}}{j}\right)
^{2}\right\}  \nonumber \label{eq45}
\end{eqnarray}
This almost linear dependence of $E$ on $j$ indicates that the
resistive mechanism of bundle activation follows Ohm law. We can
also calculate the rate of flux creep due to the thermal
activation of vortices. To do this consider hollow cylindrical
sample of a radius $r$ and the wall thickness $l<<r$ placed in the
magnetic field $ B_{ex}>B_{c1} $ applied parallel to the cylinder
axis. The sample has the trapped field $B_{in} $ inside the hole
and corresponding trapped flux $ \Phi =(B_{in}-B_{ex})\pi r^{2}$.
According to the Faraday's law electric field due to the change of
the trapped flux is equal to $(\mu _{0}/2)lr(dj/dt)$. Combining
this result with Eq.~(\ref{eq44}) we have finally
\begin{equation}
BPLdS_{c}+\ \frac{1}{2}\mu _{0}lr\frac{dj}{dt}=0 \label{eq46}
\end{equation}
This equation can be solved analytically only in the case of the
weak currents. Consequently for excitations in the form of bundle
of vortices the above equation is written as
\begin{equation}
\Omega j\exp \left\{ -\frac{\delta E_{a,b}}{k_{b}T}\left(
\frac{j_{0a,b}}{j} \right) ^{2}\right\} +\frac{dj}{dt}=0
\label{eq47}
\end{equation}
where $\Omega =\varphi _{0}\alpha _{0}j_{0}^{2}/(\mu _{0}\gamma )$
and $ \gamma =rl/(Ld)$ is the factor determined by the geometry of
the sample. The solution of Eq~(\ref{eq47}) is given in terms of
exponential integrals and for the case of $j_{0a,b}/j\ -1<<1\ $it
can be approximated as:
\begin{equation}
\frac{j(0)}{j(t)}-1=\frac{\Phi (0)}{\Phi
(t)}-1=\frac{k_{B}T}{2\delta E_{a,b} }\left(
\frac{j(0)}{j_{0a,b}}\right) ^{2}\ln \left( 1+\omega
_{a,b}t\right) \label{eq48}
\end{equation}
where
\[
\omega _{a,b}=\frac{4\varphi _{0}\alpha _{0}\delta E_{a,b}}{\mu
_{0}\gamma } \left( \frac{j_{0a,b}}{j(0)}\right) \exp \left\{
-\frac{\delta E_{a,b}}{ k_{b}T}\left( \frac{j_{0a,b}}{j(0)}\right)
^{2}\right\}
\]
This result is in agreement with the experiments on HTS.\cite{Blatter} 
For $ 0<<t<<1/\omega $ the change of trapped flux
is linear in time and for $t>>1/\omega $ logarithmic. In
antiferromagnetic superconductors, however, we see additional
change of characteristic frequency as the magnetic field changes
its direction in the $ab$ plane.
\subsection{Quantum creep}
Below $T_{0}$, the quantum activation probability is essentially
independent of temperature $P \sim \exp \left( -S/\hbar \right) $
and is interpreted as arising from the quantum tunneling of
vortices through intrinsic pinning potential.\cite{Ivlev91,Gaber95} 
In the folowing we show a considerable change of tunneling rate and 
crossover temperature due to the SF phase transition around the vortex core. 
Now, consider the vortex line as a straight string-like object of an
effective mass $m$ per unit length trapped into a metastable state
in an intrinsic pinning potential $V(u)$ and exposed to continuous
deformation $u(x,t)$ in the $\hat{z}$ direction.\cite{Krzy2000} The
magnetic field is applied in $\hat{x}$ direction ($a$ direction on
Fig.~\ref{fig3}). In the semi-classical approximation the quantum
decay rate is calculated as a saddle-point solution (bounce) of
the Euclidean action $S$ for the string
\begin{eqnarray}
S = \int\limits_{-\infty }^\infty dx \int\limits_0^{\hbar \beta
}d\tau \left\{ \frac{1}{2} m \left( \frac{{\partial u}}{{\partial
\tau }} \right)^2  + \frac{{\varepsilon _1 }}{2}\left(
\frac{\partial u}{\partial x} \right)^2  + V\left(u\right)\right.+ \nonumber \\
\left. +\frac{\eta }{2\pi }\int\limits_0^{\hbar \beta }{d\tau ^{'}
\left| \frac{u(x,\tau) - u(x,\tau ^{'} )}{\tau  - \tau ^{'} }
\right|^2 } \right\} \label{eq49}
\end{eqnarray}
Here $\beta =\left( k_{B}T\right) ^{-1}$ , $\eta $ is the
viscosity coefficient and $\tau $ denotes imaginary time. The
pinning potential $V(u)$ consists of intrinsic periodic part and
the Lorentz potential:
\begin{equation}
V\left( u\right) =-\frac{\varphi _{0}j_{0}d}{2\pi }\cos \left(
\frac{2\pi u}{ d}\right) -\varphi _{0}ju. \label{eq50}
\end{equation}
For large current, this potential can be expanded around the
inflection point to give
\begin{equation}
V\left( u\right) =V_{0}\left[ \left( \frac{u}{w}\right)
^{2}-\left( \frac{u}{ w}\right) ^{3}\right], \label{eq51}
\end{equation}
where $V_{0}=\frac{2}{3}\frac{\varphi _{0}j_{0}^{2}\pi
^{2}}{d^{2}}w^{3}$~and~$w=\frac{3d}{\pi }\left(
\frac{j_{0}-j}{2j_{0}}\right)^{\frac{1}{2}}$~may be thought as the
width of the barrier because $V(0)=V(w)=0$. The last term in
(\ref{eq49}) is the so-called Caldeira-Leggett action,
\cite{Caldeira83} which describes ohmic damping produced by the
coupling to the heat-bath of harmonic oscillators. The line
tension $\varepsilon _{1}$ is different for vortices in two
different orientations in the $ab$ plane. As previously discussed
the vortices lying parallel to $b$ direction and those laying in
the $a$ direction but created in the magnetic field fulfilling
relation $H_{c1}<H<\frac{1}{2}H_{T}$ \cite{Krzy94} have the line
tension equal to
\begin{equation}
\varepsilon _{b}=\epsilon _{0}\ln
\frac{\lambda_{ab}}{d},\label{eq52}
\end{equation}
For those vortices lying in the $a$ direction but possessing spin
flop domain, we write the following expression\cite{Krzy94}
\begin{equation}
\varepsilon _{a}=\frac{\varphi _{0}H_{T}}{2}+\frac{9}{128}\epsilon
_{0}\ln \frac{\varphi _{0}}{\pi r_{j}^{2}B_{T}}. \label{eq53}
\end{equation}
In the semiclassical approximation the decay rate is given by the
value of the action on a classical trajectory obtained from the
Euler-Lagrange equations of the motion
\begin{equation}
-m\frac{\partial ^{2}u}{\partial \tau ^{2}}-\varepsilon
_{1}\frac{\partial ^{2}u}{\partial x^{2}}+V^{^{\prime
}}(u)+\frac{\eta }{\hbar \beta } \int_{0}^{\hbar \beta }d\tau
\frac{\partial u}{\partial \tau }\cot \frac{\pi }{\hbar \beta
}\left( \tau -\tau ^{^{\prime }}\right) =0\label{eq54}
\end{equation}
The trajectory $u_{0}(x)$ for static solution of Eq.~(\ref{eq54})
gives the activation energy in the thermal regime $T>T_{0}$. Below
this crossover temperature a new kind of trajectory, periodic in
imaginary time, develops. Therefore, $u(x,\tau )$ can be expanded
in the Fourier series with Matsubara frequencies
\begin{equation}
u(x,\tau )=\sum_{n=0}^{\infty }u_{n}\left( x\right) \cos \left(
\omega _{n}\tau \right) \ \ \ ;\ \ \ \omega _{n}=\frac{2\pi
n}{\hbar \beta }. \label{eq55}
\end{equation}
Substituting this expansion into (\ref{eq54}) and linearizing
potential around the static solution $u_{0}(x)$ one obtains
\begin{equation}
-\varepsilon _{1}\frac{\partial ^{2}u_{n}}{\partial
x^{2}}+V^{^{\prime \prime }}(u_{0})u_{n}=-\left( \eta \omega
_{n}-m\omega _{n}^{2}\right) u_{n}. \label{eq56}
\end{equation}
The above equation has three discrete solutions,\cite{Landau}
the unstable one corresponding to the tunnelling process
determines the crossover temperature
\begin{equation}
k_{B}T_{0}=\frac{\hbar \eta }{4\pi m}\left\{ \left[ 1+\frac{20\pi
\varphi _{0}j_{0}m}{d\eta ^{2}}\left(
\frac{j_{0}-j}{2j_{0}}\right) ^{\frac{1}{2}} \right]
^{\frac{1}{2}}-1\right\}\label{eq57}
\end{equation}
The above calculations apply to both kind of vortices. The only
difference is their effective mass and viscosity coefficient. It
is possible to express these parameters as the function of
condensation energy accumulated in the vortex cores. For the
stationary flux flow the viscous force $ \eta \frac{\partial
u}{\partial t}$ is equal to Lorentz force. The electric field
generated by the moving vortex is $E=B\frac{\partial u}{\partial
t}$, so we get $E=\frac{\varphi _{0}B}{\eta }j=\rho j=\rho
_{N}\frac{B}{H_{c2}}j$ where $\rho _{N}$ is the normal phase
resistivity in the $ab$ plane and $ H_{c2}$ is the upper critical
field parallel to the layers. Finally,
\begin{equation}
\eta =\frac{\varphi _{0}H_{c2}}{\rho _{N}}=\frac{\varphi
_{0}\kappa H_{c} \sqrt{2}}{\rho _{N}}=\varepsilon
_{1}\frac{4\sqrt{3}\kappa ^{2}}{\pi \rho _{N}\ln \kappa
},\label{eq58}
\end{equation}
where $H_{c}=\frac{\varepsilon _{l}\kappa 2\sqrt{6}}{\pi \varphi
_{0}\ln \kappa }$ is calculated from the constitutive relation
$\varepsilon _{1}=H_{c1}\varphi _{0}$. The effective mass of the
vortex can be deduced from the work of Suhl.\cite{Suhl} He
derived the core contribution to the inertial mass
$m_{core}=\frac{3}{8}m_{e}\frac{\xi ^{2}H_{c}^{2}\mu _{0}}{
\epsilon _{F}}$, where $m_{e}$ denotes the mass of the electron
and $ \epsilon _{F}$ is the Fermi energy, and the electromagnetic
contribution coming from the energy of the electric field induced
by the moving flux. Simple estimation shows that this contribution
in layered superconductors is $10^{-4}$ of the core contribution.
Therefore,
\begin{equation}
m=\varepsilon _{1}^{2}\frac{9\lambda _{ab}^{2}m_{e}\mu
_{0}}{\varphi _{0}^{2}\pi ^{2}\epsilon _{F}\left( \ln \kappa
\right) ^{2}}. \label{eq59}
\end{equation}
It is possible now to relate the crossover temperature in
Eq.~(\ref{eq57})~to the line tension of the vortex
\begin{equation}
T_{0}=\alpha \varepsilon _{1}^{-1}. \label{eq60}
\end{equation}
The coefficient $\alpha $ depends on the material constants and
current intensity.
\subsection{Crossover from quantum to thermal creep}
As was already mentioned there are two types of vortex lines in
the system. The first ones, without magnetic domain, occur when
the field is applied in the $a$ or $b$ direction, but its
intensity does not exceed ${\frac12}H_{T}$. Eq.~(\ref{eq52}) gives
their line tension and the related crossover temperature 
$T_{0b}\sim \varepsilon _{b}^{-1}$. The other type,
possessing magnetic domain, occur when the field is applied in
the $a$ direction and its intensity exceeds ${\frac12}H_{T}$.
Eq.~(\ref{eq53}) gives their line tension and the crossover
temperature as $T_{0a}\sim \varepsilon _{a}^{-1}$. It is easy
to see that $\varepsilon _{a}>\varepsilon _{b}$ and therefore
$T_{0b}>T_{0a}$. The above calculations lead to the following
conclusion. It is possible to switch the creep regime at constant
temperature. To do this, one needs to change the field intensity
or simply change the field direction in the $ab$ plane.
\begin{figure}
\centerline{\includegraphics[width=\textwidth]{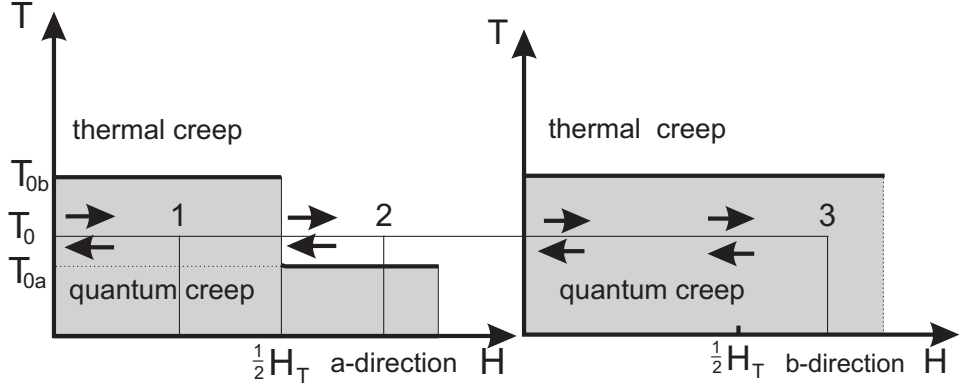}}
\caption{Schematic diagrams showing possible ways ( marked by
arrows ) of changing quantum creep behavior in the system to
thermal one, and vice versa. Shaded areas on the diagrams
correspond to quantum creep regime~\cite{Krzy2000}.} \label{fig7}
\end{figure}
The diagrams in Fig.~\ref{fig7} show possible scenarios of
crossover from quantum to thermal regimes. Let us discuss just two
of them. The first prescription is following. Fix the temperature
$T_{0}$ somewhere in the range $T_{0ba}>T_{0}>T_{0a}$. Then align
the external field in the $a$ direction and increase its intensity
to the point marked "1" on the left diagram in Fig.~\ref{fig7}.
The system is in the quantum creep regime now. Then increase the
external field beyond ${\frac12}H_{T}$. The system jumps to the
point "2" of the left diagram and finds itself in the thermal
creep regime. Doing the same operations in the reverse order one
enforces the system to crossover from thermal to quantum creep
regime. The other scenario is the following. Apply magnetic field
along $a$ axis and increase its intensity above ${\frac12}H_{T}$
keeping temperature constant in the interval
$T_{0b}>T_{0}>T_{0a}$. Then move the direction of the external
field from $a$ to $b$ axis. The system goes now from point "2" of
the left diagram (thermal creep) to the point "3" of the right
diagram (quantum creep).

\section*{ACKNOWLEDGMENTS}
Illuminating discussions with P. Tekiel, K. Rogacki and M. Ciszek
are gratefully acknowledged. This research is supported by the
State Committee for Scientific Research (KBN) within the Project
\mbox{No. 2 P03B 125 19}.


\begin{thebibliography}{99}
\bibitem{Ternary} For review see \emph{Superconductivity in Ternary Compounds},
edited by M. B. Maple, and \O. Fischer, Springer-Verlag, Berlin,
1982.
\bibitem{Kasperczyk80} J. Kasperczyk and P. Tekiel {\it Acta Phys. Polon. A} {\bf 57}, 11 (1980)
\bibitem{BulBuzdKulPanj}  L. N. Bulaevskii, A. I. Buzdin, M. Kuli\'{c} and S. V. Panjukov,
{\it Advances in Physics} {\bf 34}, 176 (1985); {\it Sov. Phys.
Uspekhi} {\bf27}, 927(1984).
\bibitem{Maple95}  M. B. Maple, {\it Physica B } {\bf 215}, 110(1995) .
\bibitem{Bauer} L. Bauernfeind, W. Widder and H. F. Braun,  {\it Physica
C} {\bf 254}, 151(1995) .
\bibitem{Pringle}   D. J. Pringle, J. L. Tallon, B. G. Walker and H. J. Trodahl,
{\it Phys. Rev. B} {\bf 59}, R11679 (1999).
\bibitem{KlamutX}  P. W. Klamut, B. Dabrowski, S. Kolesnik, M. Maxwell and J. Mais,
{\it Phys. Rev. B} {\bf 63}, 224512 (2001).
\bibitem{Houzet}  M. Houzet, A. I. Buzdin and M. Kuli\'{c}, {\it Phys. Rev.
B} {\bf 64}, 184501(2001).
\bibitem{Saxena}  S. S. Saxena, P. Agarwal, K. Ahilan, F. M. Grosche, R. K. W. Haselwimmer,
M. J. Steiner, E. Pugh, I. R. Walker, S. R. Julian, P. Monthoux,
G. G. Lonzarich, A. Huxley, I. Sheikin, D. Braithwaite and J.
Flouquet, {\it Nature} {\bf 406}, 587(2000).
\bibitem{Pfleiderer}    C. Pfleiderer, M. Uhlarz, S. M. Hayden, R. Vollmer, H. v.Lohneysen,
N. R. Bernhoeft and G. G. Lonzarich, {\it Nature} {\bf 412},
58(2001).
\bibitem{Sato}   N. K. Sato, N. Aso, K. Miyake, R. Shiina, P. Thalmeier, G. Varelogiannis,
C. Geibel, F. Steglich, P. Fulde and T. Komatsubara, {\it Nature}
{\bf 410}, 340(2001).
\bibitem{Krzy80}  T. Krzyszto\'n, {\it J. Mag. Mag. Mater.} {\bf 15-18}, 1572(1980).
\bibitem{Krzy84}  T. Krzyszto\'n, Phys. {\it Letters A} {\bf 104}, 225(1984).
\bibitem{Muto86}  H. Iwasaki, M. Ikebe and Y. Muto, {\it Phys. Rev. B} {\bf 33}, 4669
(1986).
\bibitem{Buzdin}  A. I. Buzdin, S. S. Krotov and D. A. Kuptsov, {\it Solid State
Commun.} {\bf 75}, 229(1990).
\bibitem{Wong} O. Wong, H, Umezawa and J. P. Whitehead, {\it
Physica C}{\bf 158}, 32(1989).
\bibitem{Rogacki2001}  K. Rogacki, E. Tjukanoff and S. Jaakkola,
{\it Phys. Rev. B} {\bf 64}, 094520(2001).
\bibitem{Thomlinson79}  W. Thomlinson, G. Shirane, D. E. Moncton, M. Ishikawa and \O.
Fischer, {\it J. Appl. Phys.} {\bf 50}, 1981(1979).
\bibitem{Thomlinson82}  W. Thomlinson, G. Shirane, J. W. Lynn and D. E. Moncton, in
\emph{Superconductivity in Ternary Compounds}, edited by M. B.
Maple, and \O. Fischer, Springer-Verlag, Berlin 1982.
\bibitem{Tinkham}  M. Tinkham, \emph{Introduction to Superconductivity}, chapter 5,
McGraw-Hill Inc., New York 1975.
\bibitem{Krzy2002}  T. Krzyszto\'n and K. Rogacki to be
published.
\bibitem{Lynn92}  J. W. Lynn, {\it J. Alloys and Compounds} {\bf 181 }, 419(1992).
\bibitem{Zaretsky}  J. Zaretsky, C. Stassis, A. I. Goldman, P. C. Canfield, P. Dervenagas,
B. K. Cho and D. C. Johnston, {\it Phys. Rev. B} {\bf 51},
678(1995) .
\bibitem{Lynn90}  J. W.Lynn, T. W. Clinton, W-H. Li, R. W. Erwin, J. Z .Lin, R.
N. Shelton and P.Klavins, {\it J. Appl. Phys.} {\bf 67},
4533(1990).
\bibitem {Muller} For review see  K-H. M\"uller and V. N. Narozhnyi, {\it Rep.
Prog. Phys.} {\bf 64}, 943(2001).
\bibitem{Sinha95}  S. K. Sinha, J. W. Lynn, T. E. Grigereit, Z. Hossain, L. C. Gupta,
R. Nagarajan and C. Godard, {\it Phys. Rev. B} {\bf 51},
681(1995).
\bibitem{Szymczak95}  R. Szymczak, M. Baran, L. G\l{}adczuk, H. Szymczak, Z. Drzazga
and A. Winiarska, {\it Physica C} {\bf 254},  124(1995).
\bibitem{Eskildsen2001}  M. R. Eskildsen, A. B. Abrahamsen, D. Lopez, P. L. Gammel,
D. J. Bishop, N. H Andersen, K. Mortensen and P. C. Canfield, {\it
Phys. Rev. Lett.} {\bf 86}, 320(2001).
\bibitem{Canfield98} P. C. Canfield, P. L. Gammel and D. J.
Bishop {\it Physics Today} {\bf 10}, 40(1998).
\bibitem{Krzy94}  T. Krzyszto\'{n}, {\it Phys.Lett. A} {\bf 190} (1994) 196.
\bibitem{Clem90}  J. R. Clem and M. W. Coffey, {\it Phys. Rev. B } {\bf 42},
6209(1990).
\bibitem{Kogan81}  V. G. Kogan, {\it Phys. Lett. A} {\bf 85}, 298(1981).
\bibitem{ClemLT}  J. R. Clem, in \emph{Proceedings of the 13th Conference on Low
Temperature Physics (LT 13)}, vol. 3, Plenum-Press, New York 1974,
p. 102.
\bibitem{Chakravarty90}  S. Chakravarty, B. I. Ivlev and Y. N. Ovchinnikov, {\it Phys.
Rev.B} {\bf 42}, 2143(1990).
\bibitem{Brandt92}  E. H. Brandt, {\it Physica C} {\bf 195}, 1(1992).
\bibitem{Krzy98} T. Krzyszto\'n, {\it Physica C} {\bf 294}, 47(1998).
\bibitem{Tekiel}  P. Tekiel, {\it Z. Phys. B} {\bf 104}, 423(1997).
\bibitem{Blatter}  G. Blatter, M. V. Feigelman, V. B. Geshkenbein and A. I. Larkin,
{\it Rev. Mod. Phys.} {\bf 66}, 1125(1994).
\bibitem{Ivlev91} B. I. Ivlev, Yu. N. Ovchinnikov and R. S. Thompson, {\it Phys. Rev. B} {\bf
44}, 7023(1991).
\bibitem{Gaber95} W. M. Gaber and B. N. Achar, {\it Phys. Rev. B} {\bf 52}, 1314(1995),
\bibitem{Krzy2000} T. Krzyszto\'n, {\it Physica C} {\bf 340},
156(2000).
\bibitem{Caldeira83} A. O. Caldeira and A. J. Leggett, {\it Ann.Phys. (N.Y)} {\bf 149},
374(1983).
\bibitem{Landau} L. D. Landau and E. M. Lifshitz, \emph{Quantum Mechanics}, Oxford, Pergamon
Press 1962
\bibitem{Suhl} H. Suhl, {\it Phys. Rev. Lett.} {\bf 14}, 226(1965).
\end{thebibliography}
\end{document}